\documentclass[conference]{IEEEtran}
\IEEEoverridecommandlockouts
\usepackage{cite}
\usepackage{amsmath,amssymb}
\usepackage{graphicx}
\usepackage{xcolor}
\usepackage{booktabs}
\usepackage{array}
\usepackage{tabularx}
\usepackage[hidelinks]{hyperref}

\newif\ifmarkup  \markupfalse         
\ifmarkup
  \newcommand{\todo}[1]{\textcolor{red}{\textbf{[TODO: #1]}}}
\else
  \newcommand{\todo}[1]{}
\fi

\begin{document}

\title{Mapping E-textiles Design Pain Points and Generative AI Opportunities:
Insights from Workshops in Shanghai and Winchester}

\author{
  \IEEEauthorblockN{
    Zhuchenyang Liu\IEEEauthorrefmark{1}, 
    Nianchong Qu\IEEEauthorrefmark{2}, 
    Yao Zhang\IEEEauthorrefmark{1}, 
    Marie O'Mahony\IEEEauthorrefmark{3}, 
    Qi Wang\IEEEauthorrefmark{2}, and 
    Yu Xiao\IEEEauthorrefmark{1}
  }
  \IEEEauthorblockA{
    \IEEEauthorrefmark{1}Aalto University, Espoo, Finland
  }
  \IEEEauthorblockA{
    \IEEEauthorrefmark{2}Tongji University, Shanghai, China
  }
  \IEEEauthorblockA{
    \IEEEauthorrefmark{3}University of Southampton, Southampton, UK
  }
}

\maketitle

\begin{abstract}
E-textile design involves complex decisions across materials, sensor and actuator structures, fabrication, garment integration, and data processing. It typically requires iterative prototyping and testing, which are time- and labour-intensive, while few practitioners possess cross-disciplinary expertise across all relevant domains. To identify current bottlenecks and explore how Generative AI (GenAI) might support the design process, we conducted two half-day co-design workshops, one in Shanghai and one in Winchester, with practitioners from materials science, electronics, garment design, human–computer interaction, and manufacturing. Twenty practitioners participated in the Shanghai workshop; ten of them had prior experience in e-textiles and form the contributing sample analysed here. A further ten practitioners participated in the Winchester workshop.

Participants mapped their own design pipelines, annotated bottlenecks, and proposed where GenAI could provide support. Rather than presenting a ranked list of opportunities, we report a process map that indexes each proposed GenAI role to the pipeline stage at which practitioners located it, together with the conditions on which they stated its usefulness would depend. Across both sites, practitioners consistently identified domain-specific operational barriers, including data scarcity, the disconnect between prototyping and manufacturing, and trade-offs in material–hardware integration. They also emphasized that the primary barrier to GenAI-driven e-textile design is not general model capability, but the lack of standardized, machine-readable representations of e-textile designs. Based on these findings, we identify four classes of domain-tailored AI tools that could support future e-textile design processes.

\end{abstract}


\section{Introduction}

E-textiles integrate electronic components such as sensors and actuators directly into
textiles, supporting functions that range from physiological monitoring and motion
capture to tactile interaction.
Designing e-textiles for specific applications
demands cross-disciplinary collaboration as well as an
iterative design process. Taking textile-based sensors as an example, design involves a series of coupled decisions, including material selection, sensor dimensions and structure, textile integration methods, sensor placement on the body, and, where applicable, the selection of machine learning models. Each decision constrains the others, and the consequences of a given choice often become apparent only after a prototype has been built and tested \cite{samper2020efficient}.

In practice, few practitioners possess cross-disciplinary expertise across all relevant domains. As a result, design typically proceeds through cross-disciplinary coordination and repeated cycles of prototyping, user testing, data collection, model training, and refinement. Because the design space is large and the coupling between design decisions is often poorly characterised, these cycles are slow and costly, and they frequently result in solutions that are merely workable rather than optimal.





Generative AI is advancing quickly in adjacent design domains, including artificial
electromagnetic material development \cite{zhao2025metagen}, metaphorical product design
\cite{zhou2025productmeta} and CAD generation \cite{schupbach2025text}. However, its application to
e-textile design remains largely unexplored. Before proposing a technical solution it is
necessary to establish what practitioners in this field actually need, and under what
conditions they would use such a tool. We therefore conducted two half-day co-design workshops, one in Shanghai
and one in Winchester, with practitioners from materials science,
electronics, garment design, human–computer interaction, and
manufacturing. 

The workshops aimed to address the following five questions: (1) how practitioners conceptualise
the e-textile design pipeline and the stakeholder roles within it; (2) which pain points arise at each stage; (3) where GenAI could add value; (4) which risks and concerns
practitioners anticipate; and (5) what is common and what differs between the two sites. 

Twenty practitioners participated in the Shanghai workshop; ten had prior experience in e-textiles or wearable design and form the contributing sample analysed here. A further ten practitioners participated in the Winchester workshop. During the workshops, participants worked in groups of five to map their own design pipelines, annotate bottlenecks, and identify where GenAI could provide support.

Based on the workshop results, this paper makes three contributions. First, we provide an empirically grounded map of proposed GenAI roles in e-textile design, linking each role to the pipeline stage at which practitioners located it and to the conditions they considered necessary for its usefulness. By foregrounding these conditions, the map moves beyond a wish list of possible AI functions and specifies the practical dependencies that would shape their deployment. Second, through cross-site analysis, we identify three  layers of domain-specific operational constraint: data scarcity, the disconnect between prototyping and manufacturing, and trade-offs in material–hardware integration. Third, we show that the primary barrier to GenAI-driven e-textile design is not perceived to be general model capability, but the absence of standardized, machine-readable representations of e-textile designs. Building on this analysis, we delineate four classes of domain-tailored AI tools for future e-textile design processes: a cross-disciplinary translation assistant, a constraint-aware material and process recommender, an manufacturable fabrication route advisor, and a pre-trial layout simulator.

\section{Related Work}
\label{sec:related}

Several studies have mapped the multidisciplinary nature of e-textile design processes and the coordination challenges they entail. Service blueprinting has been used to identify points at which knowledge fails to transfer between textile and electronics specialists \cite{van2024weaving}. Structured workshops have similarly been used to support shared vocabulary-building between designers and engineers \cite{zeagler2013electronic}. Workflow models for e-clothing development have also identified communication gaps and textile–electronics integration conflicts as recurring obstacles \cite{wang2026integrating}. Together, these studies describe how e-textile design is organised and where it tends to break down. However, they do not examine where a generative tool might intervene in the process, what roles practitioners would assign to such a tool, or what conditions would need to be in place for its use to be credible. This is the gap addressed in the present study.

Generative systems have been demonstrated in design domains adjacent to e-textiles. For example, MetaGen uses a language model to support metasurface element design  \cite{zhao2025metagen};
ProductMeta supports metaphorical product ideation \cite{zhou2025productmeta}; and LLM agents
have been used to generate CAD geometry from text \cite{schupbach2025text}. Although these systems differ in domain and implementation, they share an important underlying condition: each operates on an existing formal representation of the artefact, such as a parameter vector, design grammar, or geometric model. This condition is easily overlooked when discussing GenAI for design, but it is central to e-textiles, where design knowledge is distributed across materials, textile structures, electronics, fabrication processes, body placement, and use contexts. As we return to in Section~\ref{sec:discussion}, e-textile design currently lacks an equivalent standardized representation on which GenAI systems could reliably operate.

A separate line of work has examined how designers engage with generative AI in practice. Qualitative studies report that professional designers often require visual and multimodal inputs rather than purely textual interaction, that current tools provide limited support for iterative design development, and that adoption depends strongly on designers’ roles, tasks, and existing workflows \cite{park2024we, fang2026beyond}. These studies show that practitioner needs cannot be inferred directly from model capability: whether GenAI is useful depends on how it fits into situated design work. However, this literature has not yet addressed e-textiles, where the design process is shaped by distinctive constraints, including material properties, textile fabrication, bodily integration, and hardware–software coupling.


We have previously examined human–AI collaboration within e-textile design in a controlled study of flex sensor placement for shoulder motion detection \cite{liu2026exploring}. That study compared designs produced by large language models (LLMs) alone, by human designers alone, and by human designers working with LLMs. It showed how GenAI can influence design outcomes in a specific e-textile task. However, the study was confined to one task and one site. The present paper asks a complementary question: across the broader e-textile design pipeline, what do practitioners want GenAI tools to do, where in the workflow do they see these tools as useful, and what technical, representational, and organisational conditions would have to be met for such tools to be adopted? By answering these questions, the present study shifts the focus from demonstrating what GenAI can generate in a bounded design task to identifying where GenAI might fit within practitioners’ existing e-textile workflows and what conditions are required for its meaningful use.

\section{Materials and Methods}

\subsection{Workshop design}

We ran two half-day workshops, one in Shanghai in December 2025 and one in Winchester in April 2026. Both sites host active e-textile research
and teaching programmes with established industrial links, which gave access to
practitioners across materials, electronics, garment and fashion design, human--computer
interaction, and manufacturing. Recruitment advertisements were distributed through a
WeChat channel, the E-Textiles Network and LinkedIn. Applicants completed a pre-workshop
survey covering expertise, role, years of experience and expectations, and the responses
were used to form balanced multidisciplinary groups.

The workshops were conducted as facilitated group discussions. Each structured discussion began with pipeline mapping. Participants first mapped their own design processes individually and then merged these into a shared group pipeline on a canvas, annotating each stage with the stakeholders responsible for it. They then identified pain points through a round-robin process, in which each participant named their most pressing problem before any participant contributed a second item. Participants subsequently proposed opportunities for GenAI support, discussed associated risks, and presented a synthesis to the other group. In the Winchester workshop, participants additionally prioritised the resulting notes through dot voting. All sessions were audio-recorded with informed consent, and the data were anonymised before analysis.


\begin{table*}[t]
\caption{Participant profiles, from the pre-workshop survey. Technical expertise was a
multiple-choice question, so those counts exceed the number of participants.}
\label{tab:part}
\centering
\footnotesize
\renewcommand{\arraystretch}{1.15}
\begin{tabular}{@{}l p{5.0cm} p{2.1cm} p{3.1cm} p{2.6cm}@{}}
\toprule
\textbf{Site} & \textbf{Technical expertise} & \textbf{Affiliation} &
\textbf{Experience with e-textile design} & \textbf{Experience with AI tools}\\
\midrule
Shanghai ($n=10$) &
Textile design (6); HCI / interaction design (6); electronic engineering (1);
material science (2); AI / software (3) &
Academia (7); industry (3) &
$<$2 years (2); 2--5 years (2); 5--10 years (3); 10+ years (3) &
Yes (6); occasionally (4); no (0)\\
\addlinespace[2pt]
Winchester ($n=10$) &
Textile design (8); HCI / interaction design (4); electronic engineering (3);
material science (2); AI / software (1) &
Academia (9); industry (1) &
$<$2 years (2); 2--5 years (4); 5--10 years (1); 10+ years (3) &
Yes (6); occasionally (4); no (0)\\
\bottomrule
\end{tabular}
\end{table*}
\subsection{Participants}

Twenty practitioners attended in Shanghai, distributed across four groups. Inclusion in
the analysis was determined by a criterion set before the workshop: prior hands-on
experience of e-textile design. Two of the four Shanghai groups met this criterion, giving
a contributing sample of ten. Group 1 comprised academic
researchers, while Group 2 included both academic
researchers and industry practitioners. The remaining two groups contributed extensively to the
discussion of general generative-AI use but did not produce pipeline maps grounded in
first-hand e-textile work, and their canvases are not analysed here. Ten practitioners
attended in Winchester in two groups (A and B), all of whom met the criterion. Table~\ref{tab:part}
summarises the profiles.

\subsection{Data and analysis}

Each workshop produced three data streams: audio recordings of all sessions, the physical
mapping canvases, and facilitators' notes. Recordings were transcribed using Whisper \cite{radford2023robust} for the Shanghai workshop  and Otter.ai \cite{otter} for the Winchester workshop. 



Transcripts were analysed using the Framework Method \cite{gale2013using}, a matrix-based approach developed for applied multidisciplinary qualitative research. The method was appropriate for this study because it supports both deductive analysis, based on predefined categories, and inductive analysis of themes emerging from participants’ discussions.

After familiarisation with the transcripts and workshop artefacts, two researchers independently coded one primary elicitation session from each site. Their codebooks were reconciled in writing, and the agreed framework was then applied to the remaining recordings. The final framework comprised 94 codes across six categories: pipeline, stakeholders, pain points, GenAI opportunities, risks, and other. The sixth category captured material falling outside the five research questions, and each category also included an ``other''  code to retain passages that fitted the category but did not match an existing code. 

Coded material was then charted into a session-by-category matrix. The findings reported below were developed from this matrix and checked against the coded quotations. 
The six categories also structure the results:
pipeline and stakeholders are reported in Section~\ref{sec:pipelines}, pain points and opportunities in
Section~\ref{sec:map}, risks in Section~\ref{sec:risks}, and material from the sixth category is discussed across
Sections~\ref{sec:risks} and \ref{sec:discussion}. 

Each quotation reported in the paper was checked against the transcripts and confirmed to be a single continuous utterance by one speaker. The facilitators at each site were also active researchers in the field. Facilitator contributions were excluded from the analysis, with one exception: when a facilitator summarised the discussion that had just taken place, the summary was retained.


\section{Design Pipelines}
\label{sec:pipelines}
\begin{figure*}[t]
\centering
\includegraphics[width=\textwidth, trim=56 71 56 54, clip]{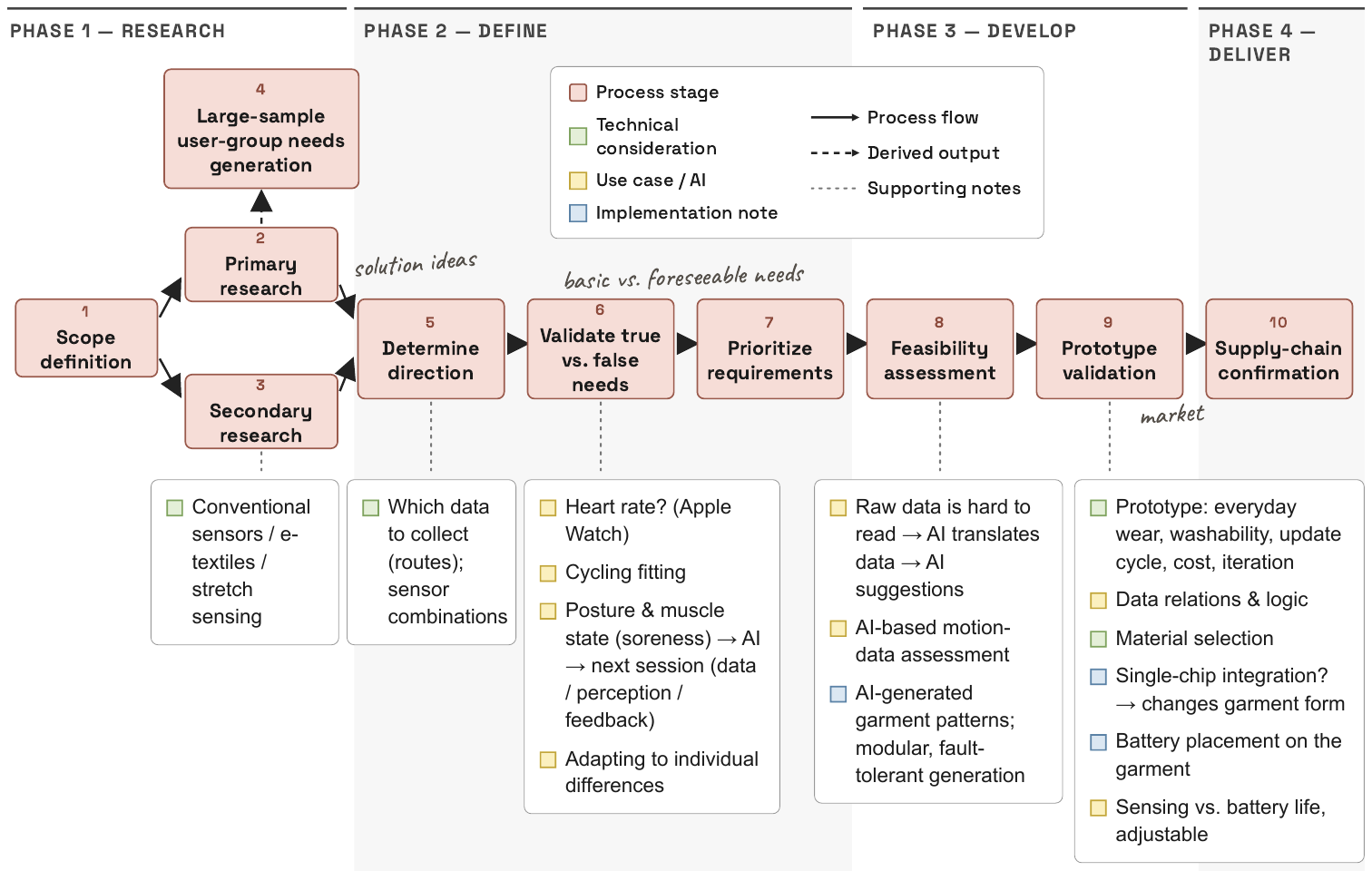}
\caption{Design pipeline mapped by Group~1 at the Shanghai workshop, redrawn from the
original canvas; notes translated by the authors. Colours follow the convention agreed
at that workshop: pink for process stages, green for technical considerations, yellow
for use cases and AI opportunities, and blue for implementation notes. Numbers give
the order in which stages were placed.}
\label{fig:sh}
\end{figure*}

\begin{figure*}[t]
\centering
\includegraphics[width=\textwidth, trim=85 56 88 53, clip]{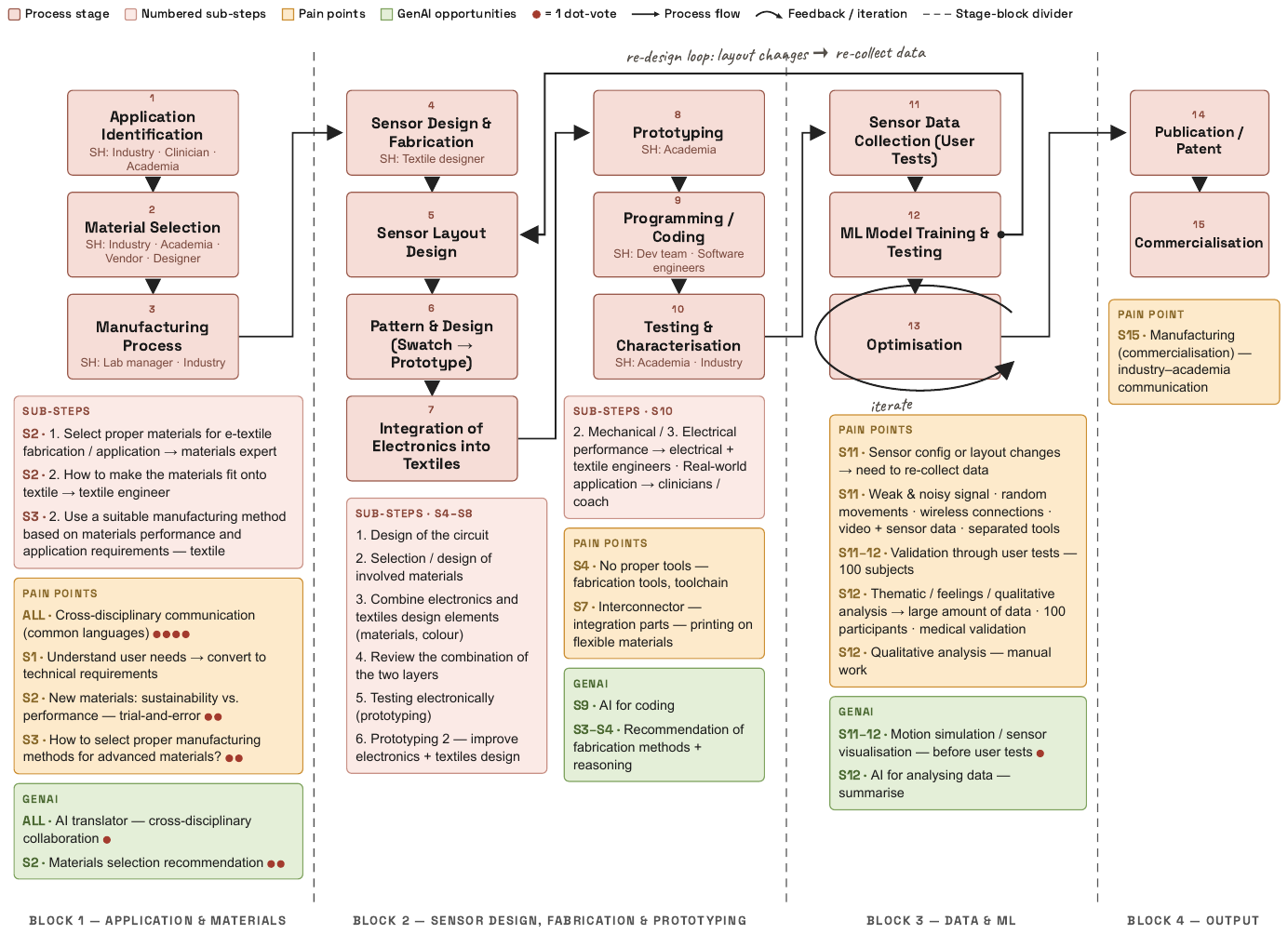}
\caption{Design pipeline mapped by Group~A at the Winchester workshop, redrawn from the
original canvas. Colours follow the convention agreed at that workshop:
pink for pipeline stages and their stakeholders, yellow for pain points, and green for
GenAI opportunities. SH denotes the stakeholders participants assigned to each stage; dots
reproduce the participants' own prioritisation vote.}
\label{fig:win}
\end{figure*}
Each group produced a canvas representing its design pipeline. Across all groups, the pipelines were consistently iterative rather than linear. We redrew two canvases for presentation: one produced by Group 1 in the Shanghai workshop (Figure \ref{fig:sh}) and one by Group A in the Winchester workshop (Figure \ref{fig:win}). These two canvases were selected because, during subsequent workshop discussions at each site, they were considered the most comprehensive representations of the design process. The two groups also had different disciplinary profiles. Group 1 in Shanghai was largely design-led, with participants mainly from product, interaction, and fashion design, whereas Group A in Winchester combined expertise in textile design, electronic engineering, and materials science.



According to the discussions within each group, practitioners defined the starting point of the e-textile design process differently depending on their roles and technical backgrounds. Product and interaction designers tended to begin by identifying user needs. Materials scientists and electronic engineers typically began with material and fabrication choices, considering fabrication implications already during material selection; materials scientists often described their contribution as ending once satisfactory electromechanical properties had been achieved in the e-textile sensors or actuators. Textile designers tended to begin with pattern making and the integration of electronic components into the garment, and their involvement often ended once the components (e.g. textile-based sensors) had been characterised or the garment was ready for data collection. Computer scientists often began by selecting ready-made sensors and determining their placement, orienting their work toward the performance of the downstream machine-learning model. This disciplinary variation is visible in the two redrawn canvases. The design-led Shanghai group tended to start from use scenarios and wearer needs, whereas the Winchester group, which included engineering expertise, gave greater prominence to materials and fabrication constraints.





By comparing Group 1 and Group 2 in the Shanghai workshop, we observed differences between academic and industry practitioners. Three participants in Group 2 worked in smart-wearable SMEs. Whereas the academic participants in Group 1 began with interviews or questionnaires to establish user needs, these industry participants relied on social-media and e-commerce platforms to identify trends and infer market demand. They also described outsourcing prototyping rather than developing prototypes in house, and reported that design decisions often rested on commercial intuition rather than formal validation. In these small firms, roles often collapsed, with one person covering several functions: one participant described having no budget to hire a researcher and therefore doing the research themselves, while hardware development was outsourced entirely. The point at which a production engineer joined the process was described as being triggered by a commercial decision rather than by technical readiness. By contrast, long-standing collaborators were said to require no explicit specification of requirements. This observation qualifies the language gap reported in Section~\ref{sec:map}: it appears most strongly in collaboration across organisations, rather than within established teams.



We did not observe a comparable academic–industry distinction in the Winchester workshop, partly because of the lower participation of industry practitioners. The two groups differed primarily in how they represented the design process: one produced a fine-grained sequence of fifteen stages with long feedback loops, while the other proposed a three-phase model in which several activities were explicitly assigned to more than one phase.


\section{Pain Points and GenAI Opportunities by Stage}
\label{sec:map}

Since practitioners may enter the process at different points, their pain points cannot be treated as a single shared list to be ranked globally. Instead, they are distributed across the pipeline according to where each discipline engages with the process. We therefore index the results by design stage rather than presenting an aggregate ranking, and report the site membership for each entry.

\begin{table*}[t]
\caption{GenAI opportunities indexed to the design stage at which practitioners located
them, with the condition each was stated to depend on. Labels PP1--PP6 and O-I--O-VII
correspond to Sections~\ref{sec:map}-A through \ref{sec:map}-G; PP7 (the requirement
that a recommendation carry its reasoning) applies across all stages as detailed in
Section~\ref{sec:map}-H. SH = Shanghai, WIN = Winchester.}
\label{tab:map}
\centering
\footnotesize
\begin{tabularx}{\textwidth}{@{}p{2.5cm} X X p{4.0cm} c@{}}
\toprule
\textbf{Stage} & \textbf{Pain point} & \textbf{Proposed role for GenAI} &
\textbf{Stated condition} & \textbf{Site}\\
\midrule
Requirement engineering &
\textbf{PP1} Converting user needs into a technical specification within a multidisciplinary project &
\textbf{O-I} Helping practitioners articulate what to ask for &
Requires domain-specific textile knowledge that is itself missing & both\\[2pt]

Material selection &
\textbf{PP2} Conductive materials transfer imperfectly to textile substrates; best-ranked material is not the best performing; sustainability sits in tension with performance &
\textbf{O-II} Recommending materials and compatible manufacturing processes &
The reasoning must be given, so that the recommendation can be evaluated against experience and communicated to others &
both\\[2pt]

Fabrication methods and manufacturability &
\textbf{PP3} Selecting a fabrication method that survives scale-up; prototypes not aligned
with bulk manufacturing facilities &
\textbf{O-III} Recommending a manufacturable fabrication method from material composition and application requirements &
Must be validated by physical evaluation; hand-feasible does not imply machine-feasible & both\\[2pt]

Integration of electronics into textiles &
\textbf{PP4} Interconnect reliability; component placement and routing disturb the garment pattern, drape, and comfort &
\textbf{O-IV} Supporting threshold decisions between competing integration routes (e.g. wires versus printed above a given length) &
Requires the objectives (reliability, manufacturability, sustainability, garment impact) to be stated explicitly & both\\[2pt]

Programming &
--- &
\textbf{O-V} Lowering the entry barrier for non-specialists &
The user must retain the competence to detect and correct errors & SH\\[2pt]

Data collection via user testing &
\textbf{PP5} A model trained at one sensor placement does not generally transfer to a new one; compounded by body variation, recruitment cost, and inconsistent user reports &
\textbf{O-VI} Simulating layouts before human-subject trials &
Simulation can falsify a configuration but cannot confirm one & both\\[2pt]

Qualitative analysis &
\textbf{PP6} Qualitative analysis of user feedback is manual and slow &
\textbf{O-VII} Transcription and summarisation, but not interpretation &
Interpretation requires human subjective judgement & WIN\\
\bottomrule

\end{tabularx}
\end{table*}

As shown in Table~\ref{tab:map}, each row gives the pain points
participants located at that stage, the role they proposed for GenAI there, and the
condition they stated that role would depend on. The final column records which sites
raised it. The conditions are the substantive part of the table: in every case
participants qualified the opportunity they had just described, and those qualifications
constrain what a usable tool would have to do. Pain points are labelled PP1--PP6 and the
corresponding opportunities O-I--O-VII; the same labels appear in the table, so that each
opportunity can be read against the difficulty it was proposed to address. 

\subsection{Requirement engineering}

\textbf{PP1: Converting user needs into a technical specification within a multidisciplinary project.}

Participants described this as a problem of not knowing what is possible, or what needs to be specified, when technical requirements depend on knowledge from another discipline. One participant gave two examples from their own projects: a fabric woven in the wrong orientation, and a set of LEDs specified without reference to colour. In both cases, missing domain knowledge produced specifications that were formally valid but practically inadequate. As another participant put it: ``\textit{What we as electronics engineers would know is obvious --- a textile designer wouldn't do it that way. But that's what you asked for. So you end up designing something that isn't actually useful, or could be done in a better way,
but that's not what you asked for.}''


The asymmetry tended to run in one direction: participants expected specificity around electronic components, but often encountered much vaguer descriptions of textiles.
As a participant at Winchester put it, ``\textit{It never fails to
astonish me --- you're talking to somebody about some interesting wearable, and you ask what's
the fabric? `Well, it's a synthetic.' No idea where it's come from. You don't get the same
thing from the electronic side --- people do tend to know what it is that they're plugging
in.}'' 


\textbf{O-I: Helping practitioners articulate what to ask for.} The proposed role for GenAI is to help practitioners express requirements in terms that another discipline can act on. Participants stressed, however, that such a system would need access to domain-specific textile knowledge, which they identified separately as currently absent.


\subsection{Material selection}

\textbf{PP2: Conductive materials transfer imperfectly to textile substrates, the obvious metric can mislead, and sustainability can sit in tension with performance.} Material selection attracted the largest number of pain points at both sites. Participants noted that existing
conductive materials are developed for films rather than for porous textile structures, making their behaviour on fabric difficult to predict; ink adhesion on fabric was also described as unreliable. 
They cautioned that ranking materials by an obvious metric, such
as resistivity, can be misleading. For example, one participant reported that, among the silver
inks they purchased, the ink with the lowest resistivity degraded most under abrasion, whereas
a lower-conductivity ink changed only slightly. In practice, then, ``\textit{the theoretical third
one on the list}'' may outperform the nominally best option. Participants also described
sustainability and performance as being in tension, with no systematic way to navigate the
trade-off.

\textbf{O-II: Recommending materials and compatible manufacturing processes.} The role proposed for GenAI is
to recommend candidate materials and compatible manufacturing processes for a given application. Participants were explicit that such recommendations would need
to include their reasoning. As one participant put
it, ``\textit{I suggest inkjet, but screen would be good because of this, and then it's up to
you\ldots\ at least you can feed that back to someone else.}'' The value of the recommendation therefore lies not only in the suggested material or process, but in the explanation that allows practitioners to evaluate it, apply their own experience, and communicate the rationale to others.


\subsection{Fabrication methods and manufacturability}

\textbf{PP3: Selecting a fabrication method that survives scale-up.} Materials that can be handled by hand-knitting or hand-weaving do not necessarily survive machine processing, as automated knitting and weaving place stringent mechanical demands on conductive and sensing yarns, which must retain their electrical properties throughout. A related problem is that prototypes are frequently not aligned with the bulk manufacturing facilities that would eventually produce them, so failures surface only at the transition to production. We return to this issue in Section~\ref{sec:discussion}.

\textbf{O-III: Recommending a manufacturable fabrication method from material composition and application requirements.} The proposed role for GenAI is to recommend fabrication methods that are compatible not only with a material's composition and intended application, but also with the requirements of
scaled or machine-based manufacturing. Participants emphasised that any such recommendation
would need to be validated through physical evaluation, since feasibility in hand fabrication
does not reliably predict feasibility in machine or scaled manufacturing.


\subsection{Integration of electronics into textiles}

\textbf{PP4: Integrating electronics into textiles without compromising reliability or garment construction.}
The integration of electronics into textiles raised two distinct problems. The first was the
reliability of interconnects between flexible textile substrates and rigid electronic
components, which participants identified as a dominant failure mode in sports applications.
The second was the way that electronic components, sensor placement, conductive routing, and
power-source location can disturb the textile or garment itself: they may alter the pattern,
constrain construction, affect drape and comfort, and limit where components can be placed.

\textbf{O-IV: Supporting threshold decisions between competing integration routes.}
Participants proposed that GenAI could help decide when one electronic-textile integration
route becomes preferable to another, for example in decisions of the form ``\textit{after a
certain length, it's better to use wires than\ldots\ printed.}'' Such support could help
compare routes such as printed conductors, wires, embroidery, connectors, or other attachment
methods against criteria including reliability, manufacturability, sustainability, and
impact on the garment. However, participants emphasised that these objectives would need to
be stated explicitly, since in current practice they are often left implicit.


\subsection{Programming}

\textbf{O-V: Lowering the entry barrier for non-specialists.} Programming was the one stage at
which participants reported already using generative tools routinely, and no separate pain
point was recorded for it. The opportunity they described is a lowering of the entry barrier:
designers without a technical background can produce working code. The condition
attached is that the user must retain enough competence to recognise when the output is
wrong and to intervene, a point taken up in Section~\ref{sec:risks}.

\subsection{Data collection via user testing}

\textbf{PP5: A model trained at one sensor placement does not generally transfer to another.} Data collection through user testing is costly when the data are needed to train and
test machine-learning models, and layout iteration multiplies that cost. A model trained
on data from one sensor placement encodes assumptions specific to that location, and
does not generally transfer to a new placement without recalibration, additional data
collection, or domain adaptation. The data already collected remain valid, for example
as a basis for comparing alternative layouts, but evaluating a revised configuration
typically requires a new round of data collection, labelling, and model training. The
Winchester group recorded this as a pain point and drew it explicitly as a return loop
from model training back to sensor layout design (Fig.~\ref{fig:win}).

Variation between bodies compounds the problem: ``\textit{Textiles are very individual. If you
start putting sensors in as well, one position or one garment might be very different for a
different person putting the same thing on.}'' In practice, this means that testing must include a larger and more diverse set of
participants, covering variation in body shape, size, and physical condition. Recruitment adds a further burden. One participant described a trial requiring on the order
of a hundred subjects, which they were only able to recruit after national broadcast coverage;
even then, the same product produced opposite reports from different users.

\textbf{O-VI: Simulating layouts before human-subject trials.} The proposed role for GenAI is to simulate sensor layouts in advance, so that configurations
that are unlikely to work can be eliminated before costly user testing begins. One participant
described this as follows: ``\textit{Based on my research in textile motion capture,
I will explore machine learning based on virtual sensors --- build virtual sensor data for
different responses from different textile sensors and combine them in a uniform platform so we
can simulate all these products in a virtual environment.}'' Participants were precise about the
limit of this role: simulation can establish what will not work, but cannot by itself establish what will.

\subsection{Qualitative analysis}

\textbf{PP6: Qualitative analysis of user feedback is manual and slow.}
At the analysis stage, participants distinguished sharply between administrative or preparatory work on qualitative data and the interpretive work of analysis.

\textbf{O-VII: Transcription and summarisation, but not interpretation.}
Participants regarded the transcription and summarisation of large volumes of study data as
delegatable to GenAI, but not the interpretation of that data. One participant noted a
``\textit{growing community that sort of rejects the use of AI to analyze data,}'' emphasising
that qualitative interpretation requires human subjective judgement, even while accepting
that a model could usefully summarise a long transcript.


\subsection{Reasoning capabilities}

\textbf{PP7: Recommendations need reasoning to be actionable.}
A condition that recurred across the table, and is worth stating separately, is that
participants did not treat explanation as an optional addition to a recommendation. Rather,
they treated reasoning as what makes a recommendation actionable in a multidisciplinary
team. Without an account of why a material, layout, process, or design route has been
recommended, practitioners cannot easily check it against their own experience, defend it to
a collaborator, or understand the trade-offs it entails.

This requirement was voiced as a wish for decision support that makes the consequences of a
choice visible: ``\textit{It's worth trying to use the AI to find a path-aware decision, so that
if I'm going this route, these are the things I'm missing out from leaving this path. Or
trying to find a combination solution that works best for design.}'' Here, the value of AI is
not simply that it selects an option, but that it exposes the alternatives, omissions, and
trade-offs associated with that option. As the facilitator summarised at the close of the
session, ``\textit{AI can work as a translator among the team members, and help us process
things --- like this workshop, we record it and use AI. It can recommend material layout. It
can already write software for us. Those are mainly to help make suggestions and then
optimise.}''

The two sites grounded trust in such recommendations differently. At Winchester, trust rested
on traceability: a recommendation that carries its reasoning can be checked against
practitioners' experience and communicated across disciplinary boundaries. At Shanghai, one
participant grounded trust instead in the perceived absence of a commercial motive,
contrasting a model with a human adviser whose advice might be shaped by what that adviser
has to sell.



\section{Risks and the Conditions on Trust}
\label{sec:risks}


At each site, the session that produced the most GenAI opportunities also produced the most risks. In several cases, the same speaker articulated both in the
same exchange. The
sessions did not divide into optimists and sceptics. This matters for how the
results should be read: the qualifications participants attached to each opportunity are
not objections raised by a dissenting minority, but conditions set by the same people who
proposed the opportunity. Trust in GenAI was therefore conditional rather than absent.

A recurring concern was the fabrication of plausible content. Participants were less concerned
with how often such errors occur than with when their costs become visible. A model may produce
a plausible but incorrect recommendation early in a process, but the error may not be detected
until substantial design, fabrication, or analysis work has already been built on it. At that
point, the cost is not simply correction, but rework. Participants therefore distinguished
GenAI failure from the failure of a conventional tool: the problem is not only that the tool can
be wrong, but that it can be wrong in a way that remains usable-looking long enough to become
embedded in subsequent work.


Participants at Shanghai drew an explicit boundary around what they were willing to delegate.
Repetitive work and the organisation of large volumes of information were regarded as safe to
hand over; the parts of a project they cared most about were deliberately withheld. Effective use of the tool was itself described as a skill. A problem has to be
reformulated in terms that a model can act on, and this requires the user to understand the
problem more precisely, not less. One participant characterised the resulting bottleneck as a
shortage of expert time rather than a shortage of computational capacity: generation is cheap,
but checking what has been generated remains expensive. In this account, GenAI does not simply
lower the expertise required to participate in a project. It can raise the level of expertise
required, because practitioners must be able to formulate, evaluate, and correct the outputs it
produces.

The second concern was skill dependency, and its direction is worth stating precisely.
Participants were concerned less with whether an AI-assisted novice can match an
experienced designer than with what happens when the tool is withdrawn. As one put it,
``\textit{if you took the AI away and you still needed to perform the procedure, the junior isn't
going to have any backup.}'' This is the qualitative counterpart of a result we have
reported elsewhere \cite{liu2026exploring}: in a controlled sensor-placement task, the least
experienced designer improved steadily through collaboration with a language model and
ultimately matched the best human-only outcome, while the most experienced designer's
performance declined. The two findings describe the same phenomenon from opposite ends. A
tool that most benefits those with least experience is also the tool whose withdrawal
they are least equipped to absorb.

Intellectual property and data ownership were discussed at length at Winchester. The
immediate concern was that proprietary or unpublished material entered into a commercial
model becomes retrievable by others: ``\textit{I read something where someone said they uploaded their
PhD thesis into ChatGPT and asked questions, and it was giving answers from their PhD back.
That information is now accessible to other people who ask ChatGPT this question, without them
publishing it yet.}'' Whether or
not this example reflects the technical operation of a particular system, it captures the
participants' concern: once material is submitted to a model, its future circulation may feel
difficult to control.

A related but more difficult question concerned authorship. When a design emerges from an
extended exchange with a model, participants questioned where originality begins and how
ownership should be assigned. One participant noted that ``\textit{in science, you don't really
have an IP\ldots\ it's a discovery,}'' pointing to the difficulty of applying familiar
ownership models to AI-mediated research and design work. This topic was introduced by a
facilitator rather than arising spontaneously, and its concentration at one site may therefore
reflect the structure of the session rather than a substantive difference between the two
communities.

\section{Discussion}
\label{sec:discussion}
This section synthesizes the empirical observations from both sites across three distinct operational layers of the e-textile pipeline: data availability, prototyping-to-manufacturing disconnect, and electronic-textile integration. Across these layers, concrete barriers converge to reveal a foundational technical constraint, specifically the absence of a standardized, machine-readable representation of e-textile artefacts, processes, and performance.

Both sites identify data scarcity as a primary operational barrier to applying generative AI in e-textile design, though it manifests differently across the two settings. In Winchester, published work reports materials, structures, and test conditions in incommensurable terms, leaving no standardized dataset from which a predictive model can learn. In Shanghai, relevant domain data exists across manufacturing chains, but remains proprietary and unshared within industrial pipelines. In both cases, generative models lack access to a structured domain corpus. This sets a practical
boundary on data-driven design assistance: even when general-purpose models are capable of
generating plausible suggestions, they do not have the organised e-textile data needed to make
those suggestions reliably domain-specific.

A recurring domain-specific bottleneck across both sites is the disconnect between laboratory-scale prototyping and industrial mass manufacturing. E-textile structures hand-assembled or hand-woven in early design phases frequently fail when transferred to automated machinery, as conductive yarns and printed lines exhibit different mechanical tolerances under continuous tension. While prototyping tools support small-scale functional verification, they rarely encode manufacturing constraints, causing designs to fail only after transitioning to bulk production pipelines. 

Participants also highlighted a fundamental trade-off inherent to electronic-textile integration: whether and when textile-integrated sensing is preferable to conventional rigid electronics. While rigid sensors are cheaper, more stable, and mass-manufacturable, e-textiles offer distinct advantages in comfort, form factor, and distributed spatial sensing on the body. 
However, because the value of a wearable often depends on its ability to
conform to individual body geometries and garment use contexts, the same properties that
motivate textile integration can also constrain standardised mass production.

Taken together, these site observations point to a foundational constraint that is independent of general AI model capabilities. The AI roles most requested by practitioners---such as cross-disciplinary translation and material-process recommendation---cannot function without a shared, machine-readable description of e-textile artifacts that a machine can inspect, compare, and reason over. Unlike adjacent fields like metasurface design or CAD generation, which operate on standardized parameter vectors or geometric models, e-textile design lacks a unified formal language spanning materials, textile structures,
fabrication routes, electronic components, interconnects, garment construction, sensing
performance, and test conditions. 
Constructing such a machine-readable representation is
therefore not a peripheral infrastructure task, but a prerequisite for robust AI support in
this domain.

Setting aside generic LLM utilities (such as code generation in O-V and transcription in O-VII), we synthesize the e-textile-specific opportunities (O-I--O-IV, O-VI) into four concrete classes of domain-tailored tools:

\textbf{(1)~Cross-disciplinary translation assistant} 
This tool would help practitioners express requirements in terms that collaborators from
other disciplines can act on, for example by translating between textile, electronics,
materials, and manufacturing vocabularies. Its boundary condition is the need for a shared,
machine-readable description schema; without such a schema, the system can only paraphrase
language rather than ground requirements in domain-specific constraints.

\textbf{(2)~Constraint-aware material and process recommender} 
This tool would recommend candidate materials, compatible fabrication methods, and associated
process routes for a given application. Its value would lie not only in ranking options, but
in making visible the trade-offs among conductivity, durability, adhesion, abrasion
resistance, washability, sustainability, manufacturability, and cost. Its boundary condition
is traceable and inspectable reasoning: practitioners need to understand why a recommendation
has been made in order to check it against experience and communicate it to collaborators.

\textbf{(3) Manufacturable fabrication route advisor.}
This tool would help identify fabrication methods that are not only feasible at prototype
scale, but also compatible with machine-based or scaled manufacturing. It would connect
material composition and application requirements to constraints such as machine tension,
yarn behaviour, print resolution, alignment, throughput, and production repeatability. Its
boundary condition is physical validation: feasibility in hand fabrication cannot be assumed
to predict feasibility in industrial production.

\textbf{(4)~Pre-trial layout simulator} 
This tool would simulate sensor layouts, routing strategies, and garment placements before
human-subject trials, with the aim of eliminating configurations that are unlikely to work.
Its boundary condition is epistemic: simulation can help falsify poor configurations and
reduce unnecessary physical trials, but it cannot definitively confirm behaviour on diverse
bodies, garments, and use conditions. Final validation must therefore remain empirical.



\section{Limitations}

Independent double-coding covered two of the thirteen recordings; the remaining recordings
were indexed by a single researcher. Restricting the analysis to practitioners with prior
e-textile experience is likely to under-represent the difficulties encountered by newcomers
to the field. The sample also included a limited number of industrial participants, which
means that perspectives from commercial production, supply chains, certification, procurement,
and large-scale manufacturing may be under-represented.


The closing session at Winchester took place after the Shanghai findings
had been presented to participants, and therefore cannot serve as evidence of independent
convergence between the sites. One facilitator at each site was also an active researcher in
the field and at several points contributed substantively rather than only managing the
session. Those contributions were excluded, except where a facilitator summarised the
preceding discussion; such summaries are attributed to the facilitator and are not offered as
independent evidence of participant opinion. 

Automatic transcription introduced errors that were
corrected against the audio only for quoted material. Finally, several of the differences
we observe between sites---for example, the concentration of intellectual property discussion at
Winchester and the near-absence of washing as a topic in Shanghai---may reflect the structure
of the sessions rather than a difference in position, and we report them as observations
rather than as findings.

\section{Conclusion and Future Work}

We mapped the e-textile design process with practitioners at two sites and indexed the
opportunities they identified for generative AI to the stages at which they located them.
Practitioners at both sites converge closely on what such a tool should do, and attach
conditions to every role they propose. Practitioners at both sites converge on primary operational constraints across data availability, prototyping-to-manufacturing disconnect, and electronic-textile integration. 

Our findings suggest that the foundational barrier is not the capability of current AI models, but the absence of a shared, machine-readable representation of e-textile designs. On this basis, we delineate four classes of domain-tailored AI tools designed to address workflow pain points and support future e-textile development.

Our next step follows from that conclusion. We are developing a description schema for
e-textile designs covering material, structure, fabrication route, placement and test
conditions. We intend to build the constraint-aware recommender and feasibility checker described in Section~\ref{sec:discussion} on top of this schema, and to evaluate both with practitioners at the two sites reported here.

\section*{Acknowledgments}
We would like to thank all the participants in the workshops. \todo{funding}

\section*{Generative AI Declaration}
Otter.ai and Whisper were used to transcribe the audio recordings. A large language model
was used to prepare and organise the corpus and to assist with language editing. All
coding decisions, theme development and interpretation were made and verified by the
authors.

\bibliographystyle{IEEEtran}
\bibliography{refs}

\end{document}